\documentstyle[12pt]{article}

\input epsf.tex
\newcommand{\be}{\begin{equation}}
\newcommand{\ee}{\end{equation}}
\makeatletter
\@addtoreset{equation}{section}
\makeatother
\def\thesection   {\arabic{section}}

\renewcommand{\theequation}{\thesection.\arabic{equation}}

\newcommand{\complex}{{\kern .1em {\raise .47ex
\hbox {$\scriptscriptstyle |$}}
    \kern -.4em {\rm C}}}
\newcommand{\real}{{{\rm I} \kern -.19em {\rm R}}}
\newcommand{\rational}{{\kern .1em {\raise .47ex
\hbox{$\scripscriptstyle |$}}
    \kern -.35em {\rm Q}}}
\renewcommand{\natural}{{\vrule height 1.6ex width .05em depth 0ex \kern -.35em
{\rm N}}}
\def\bea{\begin{eqnarray}}
\def\eea{\end{eqnarray}}
\def\bb#1{\hbox{\mybb#1}}
\def\nn{\nonumber}
\def\complex{\bb{C}}

\def\real{\bb{R}}
\def\rational{\bb{Q}}
\def\R4{\bb{R}^4}
\def\unita{{1 \kern-.30em 1}}

\def\f{\varphi}
\def\n{\nabla}
\def\o{\omega}
\def\O{\Omega}
\def\p{\partial}
\def\a{\alpha}
\def\b{\beta}
\def\th{\theta}
\def\s{\sigma}
\def\g{\gamma}

\def\d{\delta}

\begin{document}
\begin{titlepage}
\begin{flushright} {ROM2F/98/11}\\
\end{flushright}
\vskip 1mm
\begin{center}  
{\large \bf Gravitational Wave Radiation from Compact} \\ 
{\large \bf Binary Systems in the Jordan-Brans-Dicke Theory } \\   
\vspace{2.0cm}
{\bf M. Brunetti}\footnote{  {\sl  Dipartimento di Fisica, Universit\`a di Roma
II ``Tor Vergata"}}
\hspace{.1mm}\footnote{{\sl  I.N.F.N. Sezione di Roma II,  Via Della Ricerca
Scientifica, 00133 Roma, ITALY}}, 
{\bf E. Coccia}\hspace{1mm}$^{1\hspace{1mm}2}$,
{\bf V. Fafone}\footnote{  {\sl  I.N.F.N. Laboratori Nazionali di Frascati,
Via E. Fermi 40, 00044 Frascati (Roma), ITALY}},
{\bf F. Fucito}\hspace{1mm}$^{2\hspace{1mm}}$, 
\vskip 3.0cm
{\large \bf Abstract}\\
\end{center} 
In this paper we analyze the signal emitted by a 
compact binary system in the Jordan-Brans-Dicke theory.
We compute the scalar and tensor components of the power radiated 
by the source and study the scalar waveform. 
Eventually we consider the
detectability of the scalar component of the radiation by interferometers
and resonant-mass detectors.
\vfill
\end{titlepage}
\addtolength{\baselineskip}{0.3\baselineskip}

\section{Introduction} 
The detection of gravitational waves (GW) is a field of active research
from the point of view both of the development of suitable detectors and
of the study of possible sources and signal analysis. The detectors now 
operating as GW observatories are of the resonant-mass type and have a
sensitivity to typical millisecond  GW bursts
of $h\approx 6\times 10^{-19}$ ($h$ is the wave amplitude)
or, in spectral units, $10^{-21}$ (Hz)$^{-1/2}$ over a
bandwidth of a few Hz around 1 kHz  \cite{amaldi}. The first bound is appropriate for
describing the sensitivity to
gravitational collapses  while the square of the second bound represents 
the input GW spectrum that would produce a signal equal to the noise 
spectrum actually observed at the output of the detector. With this sensitivity
it is possible to monitor the strongest potential sources of GW in our galaxy
and in the local group (distances of $\approx 1$ Mpc). In order to improve
the sensitivity of these instruments, more advanced transducers and amplifiers 
are under development as well as new resonant-mass detectors of spherical shape.
Furthermore a huge effort is under way to build large laser interferometers.
It is widely believed that in the near future, sensitivities of the order
of $10^{-23}$ (Hz)$^{-1/2}$ over a bandwidth of several hundred Hz will be
attained allowing the observation of GW sources up to distances of the
order of $100$ Mpc \cite{amaldi}. It thus seems that the detection of GWs is 
highly probable at the beginning of the new millennium.
In addition to information of astrophysical interest, 
the detection of GW gives an opportunity to test the
content of the theory of gravity. In fact, it has been shown that a single 
spherical resonant-mass detector \cite{bccff}, or an array of
interferometers \cite{mshi}, have the capability to probe the spin content 
of the incoming GW.

One of the most intensively studied GW source is the inspiralling compact binary
system \cite{taylor} made of neutron stars or black holes. 
In the Newtonian regime,
the system has a clean analytic behaviour and emits a wave-form of increasing
amplitude and frequency that can sweep up to the kHz range of frequencies.
In this paper we study the radiation
emitted by this source in the framework of the Jordan-Brans-Dicke (JBD) theory.
We consider this theory to be of particular interest, since the coupling
between the scalar field and the metric has the same form of 
that of string theory, which is widely believed to give a consistent 
quantum extension of classical gravity.
Our main motivation then comes from the attemp to explore a possible 
experimental signature of string theory as already discussed in \cite{bccff}.
Furthermore the results obtained here generalize to any theory with a JBD type
coupling between matter and gravitation.

 There has been much work in this domain in the past years. Before going
to the plan of the paper, we shortly review it.
In \cite{wz} binary systems were first proposed as possible sources from
which extract more stringent bounds on $\omega_{_{BD}}$ (see (2.1) for its
definition) than those obtained from solar system data. An analysys of
spherically symmetric collapse of inhomogeneous dust was carried on in 
\cite{mshi} and later confirmed in \cite{sst}. The case of homogeneous
dust was treated in \cite{hcnn}. In \cite{ssm} a test particle around a Kerr
black hole was studied and results very similar to those of our section 5.1
for interferometers were found. In \cite{de1}
it was pointed out that deviations from general relativity can be much different
in strong and weak gravity. In \cite{de2} these deviations were parametrized
in a two dimensional space and exclusion plots were drawn out of the 
available data. Finally in \cite{no} spherical collapses were studied in a 
formalism which kept in account strong gravity effects.

The plan of the paper is the following: in Section two we describe 
the scalar and tensor GW solutions of the JBD theory. 
In Section three we compute the power emitted in tensor and scalar GW 
by a binary system. In Section four we concentrate on the scalar
waveform. In Section five, we study the interaction between the scalar
waveform and two types of earth-based detectors: interferometers and spherical 
resonant-mass detectors, giving limits for the detectability
of the signals coming from typical binary sources. Eventually, in 
Section six, we draw some conclusions.

\section{Scalar and Tensor GWs in the JBD Theory}

In the Jordan-Fierz frame, in which the scalar field 
mixes with the metric but decouples from matter,
the action reads \cite{bd} 
\bea 
S &=& S_{\rm grav}[\phi,g_{\mu\nu}]+S_{\rm m}[\psi_m,g_{\mu\nu}] \nn \\ 
&=& {c^3\over 16\pi}\int d^4x\sqrt{-g}\left[\phi R-
{\omega_{_{BD}}\over\phi}g^{\mu\nu}\partial_\mu\phi\partial_\nu\phi\right]
+{1\over c} \int d^4x L_{\rm m}[\psi_{\rm m},g_{\mu\nu}] \quad ,
\label{uno}
\eea
where $\omega_{_{BD}}$ is a dimensionless constant, whose lower bound
is fixed to be $\omega_{_{BD}}\approx 600$ by experimental data \cite{mssr},
$g_{\mu\nu}$ is the metric tensor, $\phi$ is a scalar field, and $\psi_{\rm m}$
collectively denotes the matter fields of the theory. 

As a preliminary analysis, we perform a weak field approximation around the
background given by a
Minkowskian metric and a constant expectation value for the scalar field 
\bea  
g_{\mu\nu}&=&\eta_{\mu\nu}+h_{\mu\nu} \nn \\
\f &=&\f_0+\xi \quad .
\label{due}
\eea
The standard parametrization $\f_0=2(\o_{_{BD}}+2)/G(2\o_{_{BD}}+3)$, with
$G$ the Newton constant, reproduces GR in the limit
$\omega_{_{BD}}\rightarrow\infty$, which implies $\f_0\rightarrow {1/G}$. 
Defining the new field
\be
\th_{\mu\nu} = h_{\mu\nu} - {1\over 2}\eta_{\mu\nu}h - 
\eta_{\mu\nu}{\xi\over \f_0}
\label{tre}
\ee 
where $h$ is the trace of the fluctuation $h_{\mu\nu}$, 
and choosing the gauge
\be  
\p_\mu \th^{\mu\nu}=0 
\label{quattro}
\ee 
one can write the field equations in the following form
\bea 
\label{cinque} 
\p_\a\p^\a\th_{\mu\nu} &=& -{16\pi\over \f_0}\ \tau_{\mu\nu}\\
\p_\a\p^\a\xi &=&{8\pi\over 2\o_{_{BD}}+3}\ S 
\label{sei} 
\eea 
where 
\bea \label{sette} 
\tau_{\mu\nu} &=& {1\over \f_0} (T_{\mu\nu}+t_{\mu\nu})\\
S &=& - {T\over 2(2\o_{_{BD}}+3)} 
\left(1 - {1\over 2}\th - 2 {\xi\over \f_0}\right) - 
{1\over 16\pi} \left[{1\over 2}\p_\a(\th \p^\a \xi) + 
{2\over \f_0} \p_a (\xi \p^\a \xi)\right] 
\label{otto} 
\eea 
In the equation (\ref{sette}), $T_{\mu\nu}$ is the matter stress-energy 
tensor and $t_{\mu\nu}$ is the gravitational stress-energy pseudo-tensor, 
that is a function of quadratic order in the weak gravitational 
fields $\th_{\mu\nu}$ and $\xi$. 
The reason why we have written the field equations at the quadratic order 
in $\th_{\mu\nu}$ and $\xi$ is that in this way, as we will see later, 
the expressions for $\th_{\mu\nu}$ and $\xi$ include all the terms of order 
$(v/c)^2$, where $v$ is the typical velocity of the source 
(Newtonian approximation). 
 
Let us now compute $\tau^{00}$ and $S$ at the order $(v/c)^2$. 
Introducing the Newtonian potential $U$ produced by the rest-mass 
density $\rho$ 
\be 
U(\vec x, t) = 
\int {\rho(\vec x', t)\over \mid\vec x - \vec x'\mid}\ d^3 x'
\label{nove} 
\ee 
the total pressure $p$ and the specific energy density $\Pi$ (that is the ratio of 
energy density to rest-mass density) we get 
(for a more detailed derivation, see \cite{will}): 
\bea \label{dieci} 
\tau^{00} &=& {1\over \f_0} \rho \quad , \\ 
S &\simeq & - {T\over 2(2\o_{_{BD}}+3)} 
\left(1 - {1\over 2}\th - 2{\xi\over \f_0}\right) \nn \\ 
&=& {\rho\over 2(2\o_{_{BD}}+3}
\left(1+\Pi-3\ {p\over \rho} + {2\o_{_{BD}}+1\over \o_{_{BD}} +2}\ U\right) 
\label{undici} 
\eea 
  
Far from the source, the equations (\ref{cinque}) and (\ref{sei}) 
admit wave--like solutions, which are superpositions of terms of the form 
\bea \label{dodici}
\th_{\mu\nu}(x) &=& A_{\mu\nu}(\vec x, \o) \exp(ik^\a x_\a) + c.c. \\
\xi(x) &=& B(\vec x, \o) \exp(ik^\a x_\a) + c.c. \label{tredici}
\eea 
Without affecting the gauge condition (\ref{quattro}), one can impose
$h=-2\xi/\f_0$ (so that $\th_{\mu\nu} = h_{\mu\nu}$). 
Gauging away the superflous components, one can write the amplitude
$A_{\mu\nu}$ in terms of the three degrees of freedom corrisponding 
to states with helicities $\pm 2$ and 0 \cite{lee}. 
For a wave travelling in the $z$-direction, one thus obtains
\be   
A_{\mu\nu}=\pmatrix{0&0&0&0\cr 0&e_{11}-b&e_{12} &0\cr 0&e_{12}
&-e_{11}-b&0\cr 0&0&0&0\cr},
\label{quattordici}
\ee 
where $b=B/\f_0$.

\section{Power emitted in GWs}

The power emitted by a source in GWs depends on the stress-energy 
pseudo-tensor $t^{\mu\nu}$ according to the following expression 
\be \label{quindici}
P_{GW} = r^2 \int \Phi d\O = r^2 \int <t^{0k}>\ \hat x_k\ d\O  
\ee 
where $r$ is the radius of a sphere which contains the source, 
$\O$ is the solid angle, $\Phi$ is the energy flux and the symbol 
$<...>$ implies an 
average over a region of size much larger than the wavelength 
of the GW. At the quadratic order in the weak fields we find
\be   
<t_{0z}> = -\hat z{\f_0 c^4\over 32\pi}
\left[{4(\o_{_{BD}}+1)\over \f_0^2}<(\p_0\xi)(\p_0\xi)> + 
<(\p_0 h_{\a\b})(\p_0 h^{\a\b})>\right]. 
\label{sedici}
\ee
Substituting (\ref{dodici}), (\ref{tredici}) into (\ref{sedici}), one gets 
\be \label{diciassette}  
<t_{0z}> = -\hat z {\f_0 c^4 \o^2 \over 16\pi}
\left[{2(2\o_{_{BD}}+3)\over \f_0^2} \mid B\mid^2 + 
A^{\a\b *}A_{\a\b}-{1\over 2}\mid {A^\a}_\a \mid^2\right],
\ee 
and using (\ref{quattordici})
\be \label{diciotto}  
<t_{0z}> = -\hat z {\f_0 c^4 \o^2\over 8\pi}
\left[\mid e_{11}\mid^2+\mid e_{12}\mid^2 + 
({2\o_{_{BD}}+3})\mid b \mid^2\right].
\ee 
From (\ref{diciotto}) we see that the purely scalar contribution,
associated to $b$, and the traceless tensorial contribution, associated to
$e_{\mu\nu}$, are completely decoupled and can thus be 
treated independently.

\subsection{Power emitted in tensor GWs} 

Eq. (\ref{sette}) differs from the corresponding tensor field in GR 
only by a multiplicative factor. Then we can directly write the final result 
using the well-known expression for the power emitted by a system of binary 
stars in GR
\be \label{diciannove} 
(P_{ten})_n = {1\over G\f_0} (P_{GR})_n 
\ee 
If we take $\o_{_{BD}}=600$ in (\ref{diciannove}), the multiplicative factor
$1/G\f_0$ differs from one for one part in $10^3$.

In (\ref{diciannove}) $(P_{GR})_n$ is the power emitted at frequency 
$n\o_0$ (where $\o_0$ is the orbital frequency) by a system of binary 
stars according to GR \cite{peters,mathews}, averaged over one period 
of the elliptical motion and calculated in the Newtonian approximation 
\be \label{venti} 
(P_{GR})_n = {32\over 5} {G^4\over c^5} {m_1^2 m_2^2 m \over a^5} g(n; e)
\ee   
where $m_1$ and $m_2$ are the masses of the two stars, $m$ is the total mass 
$m=m_1+m_2$, $a$ is the major semi-axis and $e$ is the eccentrity of the 
ellipse. The function $g(n; e)$ depends on the Bessel functions $J_k(ne)$
\bea \label{ventuno} 
g(n; e) &=& {n^4\over 8} 
[J_n^2(ne) (e^2-2)^2/(n^2e^4) + 
4 J_n^2(ne) (1-e^2)^3/e^4 \nn \\ 
&+& J_n^2(ne)/(3n^2) + 
4 J_n(ne) J_n'(ne) (e^2-2)(1-e^2)/(ne^3) \nn \\ 
&-& 8 J_n(ne) J_n'(ne) (1-e^2)^2/(ne^3)] \nn \\ 
&+& 4 J_n'^2(ne) (1-e^2)^2/e^2 + 
4 J_n'^2(ne) (1-e^2)/(n^2e^2) 
\eea
In the last phases of the binary system evolution, the orbit 
becomes more and more circular, because the bodies radiate the most 
at their closest approach \cite{peters}. In the 
case of null eccentricity $e=0$, the function $g(n; e)$  reduces to a 
Kronecker delta, 
$g(n;e=0) = \d_{n2}$, and the tensor GW frequency is twice 
the orbital frequency. Summing over all the harmonics $n$ \cite{peters}, 
one obtains 
\be \label{ventidue} 
P_{ten} = \sum_{n=1}^\infty (P_{ten})_n = 
{32\over 5} {G^3\over \f_0 c^5} {m_1^2 m_2^2 m\over a^5} f(e) 
\ee 
where 
\be \label{ventitre} 
f(e) = {1\over (1-e^2)^{7/2}} 
\left(1 + {73\over 24} e^2 + {37\over 96} e^4\right) 
\ee 
Eqs. (\ref{venti}) and (\ref{ventidue}) are obtained 
in the approximation of point-like masses (weak self-gravity). 
For compact binary systems like PSR 1913 + 16, they can be used
upon replacing the masses, $m_1, m_2$, by the Schwarzschild masses 
of the stars \cite{blanchet}.    

\subsection{Power emitted in scalar GWs} 

We now rewrite the scalar wave solution (\ref{tredici}) in the following 
way
\be \label{ventiquattro}
\xi(\vec x, t) = \xi(\vec x, \o) e^{-i\o t} + c. c. 
\ee 
{\it In vacuo}, the spatial part of the previous solution 
(\ref{ventiquattro}) satisfies the Helmholtz equation 
\be \label{venticinque}
(\n^2+\o^2) \xi(\vec x, \o) = 0 
\ee 
The solution of (\ref{venticinque}) can be written as 
\be \label{ventisei} 
\xi(\vec x, \o) = \sum_{jm} X_{jm} h_j^{(1)}(\o r) Y_{jm}(\th, \f) 
\ee 
where $h_j^{(1)}(x)$ are the spherical Hankel functions of the first 
kind, $r$ is the distance of the source from the observer, 
$Y_{jm}(\th, \f)$ are the scalar spherical harmonics and 
the coefficients $X_{jm}$ give the amplitudes of the various 
multipoles which are present in the scalar radiation field.
Solving the inhomogeneous wave equation (\ref{sei}), we find
\be \label{ventisette}   
X_{jm} = 16\pi i \o 
\int_V j_l(\o r') Y_{lm}^*(\th, \f) S(\vec x, \o)\ dV
\ee 
where $j_l(x)$ are the spherical Bessel functions and $r'$ is a radial 
coordinate which assumes its values in the volume $V$ occupied by 
the source. 

Substituting (\ref{sedici}) in (\ref{quindici}), considering the 
expressions (\ref{ventiquattro}) and (\ref{ventisei}), and  
averaging over time, one finally obtains 
\be \label{ventotto}
P_{scal} = {(2\o_{_{BD}}+3) c^4\over 8\pi \f_0} 
\sum_{jm} \mid X_{jm}\mid^2 
\ee 
To compute the power radiated in scalar GWs, one has 
to determine the coefficients $X_{jm}$, defined in (\ref{ventisette}). 
The detailed calculations can be found in the appendix A, while 
here we only give the final results. Introducing the reduced 
mass of the binary system $\mu=m_1 m_2/m$ and the gravitational 
self-energy for the body $a$ (with $a= 1, 2$)
\be \label{ventinove} 
\O_a = - {1\over 2} \int_{V_a} {\rho(\vec x) \rho(\vec x') 
\over \mid \vec x -\vec x'\mid}\ d^3 xd^3x' 
\ee    
one can write the Fourier components with frequency $n\o_0$ in the
Newtonian approximation as (see (\ref{aquindici}), (\ref{aventitre}), (\ref{atrentuno}),
(\ref{atrentadue})) 
\be\label{trenta}
(X_{00})_n = - {16\sqrt{2\pi}\over 3} {i\o_0 \f_0\over \o_{_{BD}} +2} 
{m \mu \over a}\ n J_n(ne) \\
\ee\newpage
\bea\label{trentabis}
(X_{1\pm 1})_n &=& -\sqrt{2\pi\over 3} 
{2i{\o_0}^2 \f_0 \over \o_{_{BD}} + 2} 
\left({\O_2\over m_2} - {\O_1\over m_1}\right)\ \mu a \nn \\
&& \left[\pm J_n'(ne) - {1\over e} (1-e^2)^{1/2} J_n(ne)\right] \\   
\label{trentuno}
(X_{20})_n &=& {2\over 3} \sqrt{\pi\over 5} 
{i{\o_0}^3 \f_0 \over \o_{_{BD}} + 2} 
\mu a^2 n J_n(ne) \\ 
\label{trentadue}
(X_{2\pm 2})_n &=& \mp 2 \sqrt{\pi\over 30} 
{i{\o_0}^3 \f_0 \over \o_{_{BD}} + 2} \mu a^2 \nn \\ 
&& {1\over n} 
\{(e^2-2) J_n(ne)/(ne^2) + 2 (1-e^2) J_n'(ne)/e \nn \\ 
&\mp& 2(1-e^2)^{1/2} [(1-e^2) J_n(ne)/e^2 - J_n'(ne)/(ne)]\} 
\eea    
Substituting these expressions in (\ref{ventotto}), leads to the power 
radiated in scalar 
GWs in the $n$-th harmonic 
\be \label{trentatre}
(P_{scal})_n = P_n^{j=0} + P_n^{j=1} + P_n^{j=2} 
\ee 
where the monopole, dipole and quadrupole terms are respectively 
\bea \label{trentaquattro} 
P_n^{j=0} &=& {64\over 9(\o_{_{BD}} + 2)} 
{m^3 \mu^2 G^4 \over a^5 c^5} n^2 J_n^2(ne) \nn \\
&=& {64\over 9(\o_{_{BD}} + 2)} 
{m^3 \mu^2 G^4 \over a^5 c^5}\ m(n; e) \\
\label{trentacinque}
P_n^{j=1} &=& {4\over 3(\o_{_{BD}} +2)} 
{m^2 \mu^2 G^3 \over a^4 c^3} 
\left({\O_2\over m_2} - {\O_1\over m_1}\right)^2 \nn \\ 
&& n^2 \left[ J_n'^2(ne) + {1\over e^2} (1-e^2) J_n^2(ne)\right] \nn \\ 
&=& {4\over 3(\o_{_{BD}} +2)} 
{m^2 \mu^2 G^3 \over a^4 c^3} 
\left({\O_2\over m_2} - {\O_1\over m_1}\right)^2 d(n; e) \\ 
\label{trentasei} 
P_n^{j=2} &=&  {8\over 15(\o_{_{BD}} +2)} 
{m^3\mu^2 G^4\over a^5 c^5}\ g(n; e) 
\eea 
In figures 1, 2, 3 we plot the monopole $m(n; e)$, dipole $d(n; e)$  
and quadrupole $g(n; e)$ functions against the index $n$, for different 
values of the eccentricity $e$.\newpage
\centerline{\vbox {\epsfysize = 5.4cm \epsfbox{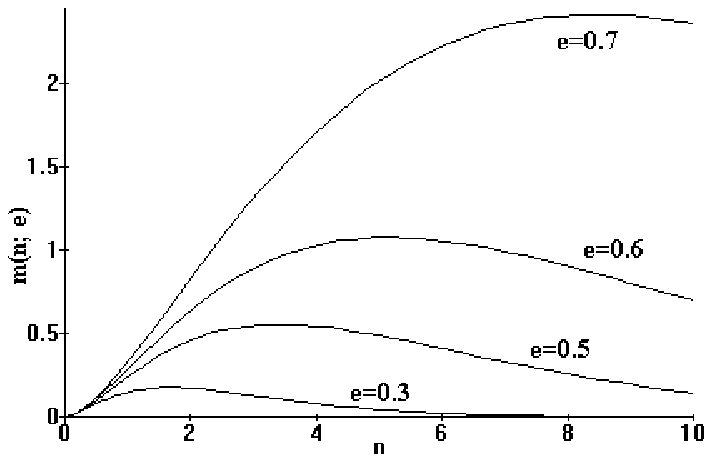}}}
\smallskip 
\centerline{\bf Figure 1}
\vskip .6cm 
\vskip .6cm 
\centerline{\vbox {\epsfysize = 5.4cm \epsfbox{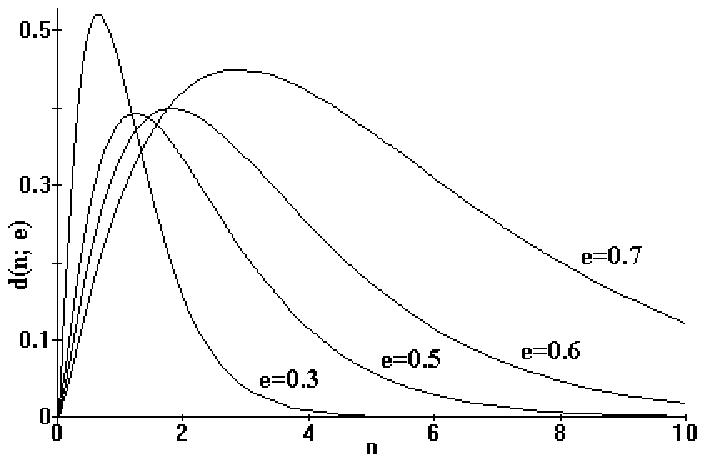}}}
\smallskip 
\centerline{\bf Figure 2}
\vskip .6cm 
\vskip .6cm 
\centerline{\vbox {\epsfysize = 5.4cm \epsfbox{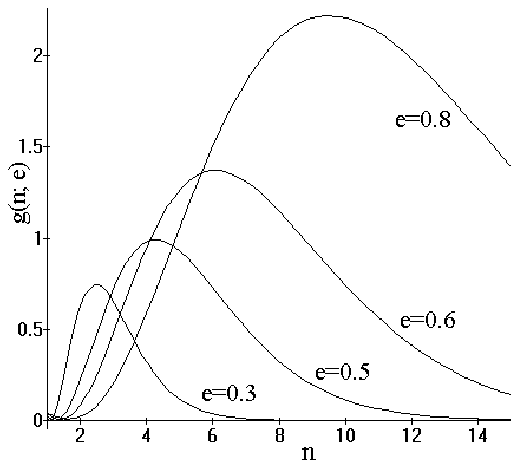}}}
\smallskip 
\centerline{\bf Figure 3}
\vskip .6cm 
From the figures one can infer the dominant harmonics in the scalar 
GWs radiation. In the case of circular orbit, 
the dipole function $d(n; e)$ reduces 
to a Kronecker delta $d(n; e=0) = \d_{n1}$, while the monopole 
function $m(n; e)$ goes to zero. 

The total power radiated in scalar GWs by a binary system is 
the sum of three terms 
\be \label{trentaseibis}
P_{scal} = P^{j=0} + P^{j=1} + P^{j=2} 
\ee 
where 
\bea \label{trentasette} 
P^{j=0} &=& {16\over 9(\o_{_{BD}}+2)} 
{G^4\over c^5} {m_1^2 m_2^2 m\over a^5} 
{e^2\over (1-e^2)^{7/2}} \left(1+{e^2\over 4}\right) \\ 
\label{trentotto} 
P^{j=1} &=& {2\over \o_{_{BD}}+2}
\left({\O_2\over m_2} - {\O_1\over m_1}\right)^2  
{G^3\over c^3} {m_1^2 m_2^2 \over a^4} 
{1\over (1-e^2)^{5/2}} \left(1+{e^2\over 2}\right) \\
\label{trentanove}  
P^{j=2} &=& {8\over 15(\o_{_{BD}}+2)} 
{G^4\over c^5} {m_1^2 m_2^2 m\over a^5} 
{1\over (1-e^2)^{7/2}} \left(1+{73\over 24} e^2 + {37\over 96} e^4\right) 
\eea 
Note that $P^{j=0}, P^{j=1}, P^{j=2}$ all  go to zero in the limit $\o_{_{BD}} 
\to \infty$.  

\section{Scalar GWs} 

We now give the explicit form of the scalar GWs
radiated by a binary system. To this end, note that 
the major semi-axis, $a$, is related to the total energy, 
$E$, of the system through the following equation
\be \label{quaranta} 
a = - {Gm_1 m_2 \over 2E} 
\ee 
Let us consider the case of a circular orbit, remembering that in the 
last phase of evolution of a binary system this condition is usually 
satisfied. Furthermore we will also assume $m_1 = m_2$.
With these positions
only the quadrupole term, (\ref{trentasei}), of the gravitational 
radiation is different from zero. 
The total power radiated in GWs, averaged over time, 
is then given by (\ref{ventidue}), (\ref{trentasette})-(\ref{trentanove})
\be \label{quarantuno} 
P = {8\over 15(\o_{_{BD}}+2)} {G^4\over c^5} {m_1^2 m_2^2 m\over d^5} 
[6(2\o_{_{BD}}+3) + 1] 
\ee
where $d$ is the relative distance between the two stars. 
The time variation of $d$ in one orbital period is 
\be \label{quarantadue} 
\dot d = - {G m_1 m_2\over 2 E^2} P
\ee 
Finally, substituting (\ref{quaranta}), (\ref{quarantuno})
in (\ref{quarantadue}) and integrating over time, one obtains 
\be \label{quarantatre}
d = 2 \left({2\over 15}  
{12\o_{_{BD}}+19\over \o_{_{BD}}+2}
{G^3 m_1 m_2 m\over c^5} \right)^{1/4} \tau^4
\ee 
where we have defined $\tau= t_c - t$, $t_c$ being the time of
the collapse between the two bodies. 

From (\ref{ventisei}), (\ref{trenta})-(\ref{trentadue}) and 
(\ref{bquattro})
one can deduce the form of the scalar field (see appendix B for 
details) which, for equal masses, is 
\be \label{quarantaquattro} 
\xi(t) = - {2 \mu\over r(2\o_{_{BD}}+3)} 
\left[v^2 + {m\over d} - (\hat n\cdot \vec v)^2 + 
{m\over d^3} (\hat n\cdot \vec d)\right] 
\ee 
where $r$ is the distance of the source from the observer, 
and $\hat n$ is the versor of the line of sight from the observer to the 
binary system center of mass. Indicating with $\g$ the inclination angle,
that is the angle between the orbital plane and the reference plane
(defined to be a plane perpendicular to the line of sight), and
with $\psi$ the true anomaly, that is the angle between $d$ and the $x$-axis
in the orbital plane $x$-$y$, yields $\hat n \cdot \vec d = d \sin\g\sin\psi$. 
Then from (\ref{quarantaquattro}) one obtains 
\be \label{quarantacinque} 
\xi(t) = {2G\mu m\over (2\o_{_{BD}}+3)c^4 d r} \sin^2\g\cos(2\psi(t))
\ee 
which can also be written as
\be \label{quarantasei} 
\xi(\tau) = \xi_0(\tau) \sin(\chi(\tau)+\bar\chi)
\ee
where $\bar\chi$ is an arbitrary phase and the amplitude $\xi_0(\tau)$ 
is given by 
\bea \label{quarantasette} 
\xi_0(\tau) &=& {2G\mu m\over (2\o_{_{BD}}+3)c^4 d r} \sin^2\g \nn \\
&=& {1\over 2(2\o_{_{BD}}+3) r} 
\left({\o_{_{BD}}+2\over 12\o_{_{BD}}+19}\right)^{1/4} 
\left({15G\over 2c^{11}}\right)^{1/4}
{{M_c}^{5/4}\over \tau^{1/4}} \sin^2\g
\eea 
In the last expression, we have introduced the definition of 
the chirp mass $M_c = (m_1m_2)^{3/5}/m^{1/5}$.

\section{Detectability of the scalar GWs}

Let us now study the interaction of the scalar GWs with two types of GW 
detectors.

As usual, we characterize the sensitivity of the detector by the spectral 
density of strain $S_h(f)$ $[\hbox{Hz}]^{-1}$. The optimum performance of a 
detector is obtained by 
filtering  the output with a filter matched to the signal. The energy 
signal-to-noise ratio $SNR$ 
of the filter output is given by the well-known formula:
\begin{equation}
SNR = \int^{+\infty}_{-\infty}\frac{|H(f)|^2} {S_{h}(f)}\,df
\label{snr}
\end{equation}
where, in our case,  $H(f)$ is the Fourier transform of the scalar 
gravitational  waveform $h_s(t)=G\/\xi_0(t)$.

We must now take into account the astrophysical
restrictions on the validity of the waveform (\ref{quarantasei}) 
which is obtained in the Newtonian approximation for point-like
masses.
In the following, we will take the point of view that this approximation
breaks down when there are five cycles remaining to collapse \cite{dewey,cf}.

The five-cycles limit will be used to restrict the range of $M_c$
over which our analysis will be performed. From (\ref{quarantatre}),
one can obtain
\bea \label{quarantotto}
\o_g(\tau) &=& 2\o_0 = 2 \sqrt{Gm\over d^3} \nn \\
&=& 2\left({15c^5\over 64G^{5/3}}\right)^{3/8}
\left({\o_{_{BD}} + 2\over 12\o_{_{BD}} + 19}\right)^{3/8}
{1\over {M_c}^{5/8}} \tau^{3/8}
\eea
Integrating (\ref{quarantotto}) yields the amount of phase
until coalescence
\be \label{quarantanove}
\chi(\tau) = {16\over 5}\left({15c^5\over 64G^{5/3}}\right)^{3/8}
\left({\o_{_{BD}} + 2\over 12\o_{_{BD}} + 19}\right)^{3/8}
\left({\tau \over {M_c}}\right)^{5/8}
\ee
Setting (\ref{quarantanove}) equal to the limit period,
$T_{5~cycles}=5(2\pi)$, solving for $\tau$
and using (\ref{quarantotto}) leads to
\be \label{cinquanta}
\o_{5~cycles} = 2\pi (6870\ \hbox{Hz})
\left({\o_{_{BD}} + 2\over 12\o_{_{BD}} + 19}\right)^{3/5}
{M_\odot\over M_c}
\ee
Taking $\o_{_{BD}} = 600$, the previous limit reads
\be \label{cinquantuno}
\o_{5~cycles} = 2\pi (1547\ \hbox{Hz})
{M_\odot\over M_c}
\ee

\subsection{Interferometers}

An interferometric detector measures the relative displacements between the 
mirrored faces of test masses arranged in the L-shaped configuration of a 
Michelson interferometer. The directivity antenna pattern for a tensorial wave 
is such that the maximum detector output is obtained for a wave inpinging 
perpendicularly respect to the plane defined by the interferometer arms.
On the contrary a scalar BD wave (which is also transverse) inpinging in the same direction will give a null effect.
In the case of a scalar wave, the maximum effect will be 
obtained for a wave propagating
along one interferometer arm. Assuming such a
direction and setting $\sin\gamma=1$ in (\ref{quarantasette}),
we can, for instance, evaluate the $SNR$ for the VIRGO 
interferometer, presently under construction. We use for $S_h(f)$ the VIRGO
noise spectrum as modeled in ~\cite{cuoco}, which is the sum of three main
components: thermal noise in the pendola, thermal noise in the mirrors, shot 
noise at high frequency:
\be \label{snrvirgo}
S_h(f) = 10^{-47} \left[\alpha_1\left(\frac{f}{100 \hbox{Hz}}\right)^{-5} +
\alpha_2\left(\frac{f}{100 \hbox{Hz}}\right)^{-1}+
\alpha_3\left(\frac{f}{100 \hbox{Hz}}\right)^{2}\right] \hbox{Hz}^{-1}
\ee
where $\alpha_1 = 2.0$, $\alpha_2 = 91.8$ and $\alpha_3 = 1.23$.

Integrating (\ref{snr}) over the range 10 Hz--500 Hz, one obtains 
the following $SNR$:
\be
SNR = 7.7 \ 10^4 \left(\frac{r}{\hbox{Mpc}}\right)^{-2} 
\left(\frac{M_c}{M_\odot}\right)^{5/3} \left(\frac{\o_{_{BD}}+2}{12\o_{_{BD}}
+19}\right)^{1/2}\left[\frac{1}{2(2\o_{_{BD}}+3)}\right]^2
\ee
For $\o_{_{BD}}=600$, we find:
\be \label{snrinteg600}
SNR = 3.8 \ 10^{-3} \left(\frac{r}{\hbox{Mpc}}\right)^{-2} 
\left(\frac{M_c}{M_\odot}\right)^{5/3}
\ee
The inspiralling of two neutron stars of 1.4 solar masses each will then give 
$SNR =1$ at a source distance $r\simeq 70$ kpc. The inspiralling of two black 
holes of 10 solar masses each will give $SNR=1$ at a distance $r\simeq 300$ kpc.
To get this last limit we have integrated (\ref{snr}) up
to $\o_{5~cycles} = 2\pi (178\ \hbox{Hz})$. 

\subsection{Spherical detectors}

A GW excites those vibrational modes of a resonant body having the proper 
simmetry. 
Most people consider the next generation of resonant-mass detectors will be of
spherical shape. In the framework of the JBD theory the spheroidal modes with 
$l=2$ and $l=0$ are sensitive to the incoming GW \cite{bccff, cola}. Thanks to its
multimode nature, a single sphere is capable of detecting GW's from all 
directions and polarizations.
We evaluate the $SNR$ of a resonant-mass detector of spherical shape for its 
quadrupole mode with $m=0$ and its monopole mode.
In a resonant-mass detector, $S_h(f)$ is a resonant curve and can
be characterized by its value at resonance $S_{h}(f_n)$
and by its half height width \cite{pizz}. $S_{h}(f_n)$ can thus
be written as
\begin{equation}
  S_h(f_n) = \frac{G}{c^3}\frac{4kT}{\sigma_{n} Q_{n} f_{n}}
\label{esseacca}
\end{equation}

Here $\sigma_{n}$ is the cross-section associated with the $n\/$-th
resonant mode,
$T$ is the thermodynamic temperature of the detector 
and $Q_n\/$ is the quality factor of the mode.

The half height width of $S_{h}(f)$ gives the bandwidth of the resonant
mode

\begin{equation}
\Delta f_n = \frac{f_n}{Q_n} \Gamma_{n} ^{-1/2}
\label{deltaeffe}
\end{equation}

Here, $\Gamma_n$ is the ratio of the wideband noise in the $n\/$-th resonance
bandwidth to the narrowband noise. 

From the resonant-mass detector viewpoint, the chirp signal can be treated as a 
transient GW, depositing energy in a time-scale short with respect to the 
detector damping time. We can then consider constant the Fourier transform of 
the waveform within the band of the detector and write \cite{pizz}
\be \label{snr0}
SNR = \frac{2\pi\Delta f_{n} |H(f_n)|^{2}}{S_{h}(f_n)}
\ee

The cross-sections associated to the vibrational modes
with $l=0$ and $l=2$, $m=0$ are respectively \cite{bbcfl}
\bea \label{cinquantaquattro}
\s_{(n0)} &=& H_n {G M {v_s}^2\over c^3 (\o_{_{BD}}+2)}
\\ 
\label{cinquantacinque}
\s_{(n2)} &=& {F_n\over 6} {G M {v_s}^2\over c^3 (\o_{_{BD}}+2)}
\eea
All parameters entering the previous equation refer to the detector
$M$ is its mass, $v_s$ the sound velocity
and the constants $H_n$ and $F_n$ are given in \cite{bbcfl}.
The signal-to noise ratio can be
calculated analytically by approximating the waveform with
a truncated Taylor expansion around $t=0$,
where $\o_g(t=0) = \o_{nl}$ \cite{clark,dewey}
\be \label{cinquantasette}
h_s(t) \approx G \xi_0(t=0)
\sin\left[\o_{nl} t + {1\over 2} \left({d\o\over dt}\right)_{t=0} t^2\right]
\ee
Using quantum limited readout systems, one finally obtains
\bea \label{cinquantanove}
(SNR_n)_{l=0} &=&
{5\cdot 2^{1/3} H_n G^{5/3}  \over
32(\o_{_{BD}} +2)(12\o_{_{BD}} + 19) \hbar c^3} \nn \\ 
&& {{M_c}^{5/3} M {v_s}^2 \over r^2 {\o_{n0}}^{4/3}}
\sin^4\g \\
\label{sessanta}
(SNR_n)_{l=2} &=&
{5\cdot 2^{1/3} F_n G^{5/3}  \over
192(\o_{_{BD}} +2)(12\o_{_{BD}} + 19) \hbar c^3} \nn \\ 
&& {{M_c}^{5/3} M {v_s}^2 \over r^2 {\o_{n0}}^{4/3}}
\sin^4\g
\eea
which are respectively the signal-to-noise ratio for the modes
with $l=0$ and $l=2$, $m=0$ of a spherical detector.

It has been proposed to realize spherical detectors with 3 meters diameter,
made of copper alloys, with mass of the order of 100 tons \cite{fc}.
This proposed detector has resonant frequencies of  $\omega_{12} = 2 \pi \cdot
807$
rad/s and $\omega_{10} = 2 \pi \cdot 1655$ rad/s.
In the case of optimally oriented orbits (inclination angle $\g = \pi/2$) and 
$\o_{_{BD}}=600$, the inspiralling of two compact objects of 1.4 solar masses 
each will then be detected with $SNR =1$ up to a source distance 
$r(\o_{10})\simeq 30$ kpc and $r(\o_{12})\simeq 30$ kpc.

\section{Summary and conclusions}

In this paper we have studied the waveforms emitted by a system of binary 
stars in the framework of the JBD theory and computed the power emitted in 
GW's for the tensor and scalar components. 
Eventually we derived limits for the detectability of such signals
by interferometers and resonant mass detectors.
In the former case we 
left aside the question of the detectability of the scalar component 
of the GW \cite{mshi} and we have concentrated on
waves impinging from the most favourable direction.
We would now like to comment
The detectability ranges 
obtained in Sections 5.1,5.2 for 
the scalar component of the GW's emitted by a binary system, vary from
few tens to few hundreds of kpc's for masses ranging from those of typical
neutron stars ($1.4 M_\odot$) to those of typical black holes ($10 M_\odot$).
We remind the reader that for the purely tensorial 
component (in this case the results obtained in the framework of the JBD theory
are practically the same of those of GR) the detectability range (for 
$1.4 M_\odot$)
is $r\simeq 120$ Mpc for spherical detectors \cite{cf} and 
$r\simeq 300$ Mpc for interferometers \cite{cuoco}.
The expected rate of coalescence events is of the order of 1 per year up
to 100 Mpc \cite{phin}. We can thus conclude that binary systems
look less promising than gravitational collapses \cite{bbcfl}
as sources of detectable scalar GW from the next generation of earth-based detectors.

\newpage

\renewcommand{\theequation}{\thesection.\arabic{equation}}

\appendix

\section{Appendix}

In order to calculate the coefficients $X_{jm}$ defined in
(\ref{ventisette}), let us first express $X_{jm}$ as a sum of Fourier 
components 
\be \label{auno} 
X_{jm}(t) = \sum_{n=-\infty}^{+\infty} (X_{jm})_n e^{in\b} 
\ee 
where $\b$ is the mean anomaly which, in terms of the orbital frequency
$\o_0$ and of the time of periastron passage $T_0$ (or equivalently
in terms of the eccentric anomaly $\a$ and the eccentricity), results 
\be \label{adue}
\b = \o_0 (t-T_0) = \a -e \sin\a
\ee
In the so-called quadrupole approximation $y= \o d/c \ll 1$, the
spherical Bessel functions $j_l(y)$, which appear in
(\ref{ventisette}), can be written as
\be \label{atre}
j_l(y) = \sum_{k=0}^\infty {(-1)^k y^{l+2k} \over 2^k k! (2l+2k+1)!!}
\ee
Making use of the Newtonian approximation (including only
two terms of the series (\ref{atre})),
from (\ref{undici}) and (\ref{ventisette}) one can obtain
$X_{00}$ as
\be \label{aquattro}
X_{00} = {4\pi i \o \f_0 \over \o_{_{BD}} +2}
\int_V \left(1- {(\o r')^2\over 6}\right) Y_{00} \rho
\left(1 + \Pi -3\ {p\over \rho} +
{2\o_{_{BD}} +1 \over \o_{_{BD}} +2}\ U\right) dV
\ee
To simplify (\ref{aquattro}) we use the post-Newtonian expressions of
the conserved quantities \cite{will}
\bea \label{acinque}
P^0 &=& \int \rho \left(1 + v^2 +
{5\o_{_{BD}} + 4\over 2(\o_{_{BD}} + 2)}\ U+ \Pi \right)d\vec x \\
P^i &=& \int \rho \left(1 + v^2 +
{5\o_{_{BD}} + 4\over 2(\o_{_{BD}} + 2)}\ U+ \Pi +
{p\over \rho}\right)v^i d\vec x \nn \\
&-& {1\over 2}\int \rho \left(1 + {v^2\over 2} +
{3(\o_{_{BD}}+1)\over \o_{_{BD}} + 2}\ U\right) W^i d\vec x
\label{asei}
\eea
where
\be \label{asette}
W^i = \int \rho' {[\vec v'\cdot \mid\vec x-\vec x'\mid](x-x')_i\over
\mid \vec x-\vec x'\mid^3}\ d\vec x'
\ee
To the required accuracy and modulo constants, one then obtains
\be \label{aotto}
X_{00} = - {\sqrt{4\pi} i \o \f_0 \mu \over \o_{_{BD}} + 2}
\left(v^2 + {(\o d)^2\over 6} + {m\over r} \right)
\ee
In terms of the eccentric anomaly $\a$, the above expression reads
\be \label{anove}
X_{00}(t) = - {\sqrt{4\pi} i \o \f_0 \mu m \over a(\o_{_{BD}} + 2)}
G_{j=0}
\ee
where
\be \label{adieci}
G_{j=0} = {1+e\cos\a\over 1-e\cos\a} + {n^2\over 6} (1-e\cos\a)^2 +
{1\over 1-e\cos\a}
\ee
Let us express $G_{j=0}$ as a sum of Fourier components
\be \label{aundici}
G_{j=0} = \sum_{n=0}^\infty A_n \cos(n\b)
\ee
where $\b$ is the mean anomaly defined in (\ref{adue}). Using the integral
expression of the Bessel functions
\be \label{adodici}
J_n(x) = {1\over \pi} \int_0^\pi \cos(n\a- x\sin\a)d\a
\ee
and the recursion formula
\be \label{atredici}
J_{n+1}(x) + J_{n-1}(x) = {2n\over x} J_n(x)
\ee
results
\be \label{aquattordici}
A_n = {2\over \pi} \int_0^\pi G_{j=0} \cos(n\b) d\b = {16\over 3} J_n(ne)
\ee
Then one finally has
\be \label{aquindici}
(X_{00})_n = - {16\sqrt{2\pi}\over 3} {i\o_0 \f_0\over \o_{_{BD}} +2}
{m \mu \over a}\ n J_n(ne)
\ee
Choosing the orbital plane as the $x$-$y$ plane, yields  $X_{10}=0$
and
\be \label{asedici}
X_{1\pm 1} = {4\pi i\o \f_0 \over \o_{_{BD}} + 2}
\int_V {\o r'\over 3}\ Y_{1\pm 1}^*(\th, \phi) \rho
\left(1+\Pi -3 {p\over \rho} + {2\o_{_{BD}} + 1\over \o_{_{BD}} + 2} U\right)
\ee
Defining ${\cal G} = \O_2/m_2 - \O_1/m_1$, one then
obtains to the required order
\be \label{adiciassette}
X_{1\pm 1} = {4\pi i\o^2 \f_0 \mu {\cal G} \over \sqrt{6\pi}(\o_{_{BD}} + 2)}
d(\cos\psi \mp i \sin\psi)
\ee
where $\psi$ is the true anomaly. In terms of the eccentric anomaly $\a$,
(\ref{adiciassette}) results
\be \label{adiciotto}
X_{1\pm 1} = - \sqrt{2\pi\over 3} {2i\o^2\ \f_0 {\cal G} \mu a
\over \o_{_{BD}} +2} G_{j=1}
\ee
where
\be \label{adiciannove}
G_{j=1} = \pm (\cos\a -e) -i (1-e^2)^{1/2} \sin\a
\ee
The binary system center of mass calculated with respect to the gravitational
self-energies $\O_a$ doesn't coincide with the center of mass with respect
to the inertial masses $m_a$ of the two bodies, if the masses are
different (Nordtvedt effect): the resulting dipole moment is, as we have seen,
a source of scalar radiation. If we express $G_{j=1}$ as a sum of Fourier
components
\be \label{aventi}
G_{j=1} = B_0 + \sum_{n=1}^\infty [B_n \cos(n\b) + C_n \sin(n\b)]
\ee
we obtain
\be \label{aventuno}
B_n = \pm {2\over n} J_n'(ne)
\ee
(where the prime indicates derivative with respect to the argument $ne$) and
\be \label{aventidue}
C_n = {2\over \pi} \int_0^\pi G_{j=1} \sin(n\b) d\b = - {2i\over ne}
(1-e^2)^{1/2} J_n(ne)
\ee
Then the $n$-th component of the coefficient $X_{1\pm 1}$ is
\bea
(X_{1\pm 1})_n &=& -\sqrt{2\pi\over 3} 
{2i{\o_0}^2 \f_0 \over \o_{_{BD}} + 2} {\cal G} \mu a \nn \\
&& \left[\pm J_n'(ne) - {1\over e} (1-e^2)^{1/2} J_n(ne)\right] 
\label{aventitre}
\eea
Finally, in the case $j=2$ one obtains $X_{2\pm 1} =0$ and
\bea \label{aventiquattro}
X_{20} &=& {4\pi i \o^3 \f_0 \over 15(\o_{_{BD}} + 2)}
\int_V r'^2 Y_{20}^*(\th, \phi) \rho dV \\
X_{2\pm 2} &=& {4\pi i \o^3 \f_0 \over 15(\o_{_{BD}} + 2)}
\int_V r'^2 Y_{2\pm 2}^*(\th, \phi) \rho dV
\label{aventicinque}
\eea
and in terms of $\a$, using the Newtonian approximation,
\bea \label{aventisei}
X_{20}(t) &=& \sqrt{\pi\over 5}
{i \o^3 \f_0 \mu a^2\over 3(\o_{_{BD}} + 2)} (1-\cos\a)^2 \\
X_{2\pm 2} &=& \mp \sqrt{\pi\over 30}
{i \o^3 \f_0 \mu a^2 \over (\o_{_{BD}} + 2)} G_{j=2} 
\label{aventisette}
\eea
where
\be \label{aventotto}
G_{j=2} = D_0 + \sum_{n=1}^\infty [D_n \cos(n\b) + E_n \sin(n\b)]
\ee
Calculating $D_n$ and $E_n$
\bea \label{aventinove}
D_n &=& - {4\over n} \left[ {1\over ne^2} (2-e^2) J_n(ne) +
{2\over e} (e^2-1) J_n'(ne)\right] \\
E_n &=& \mp {8i\over n} \left[ {1-e^2\over e^2} J_n(ne) -
{1\over ne} J_n'(ne) \right]
\label{atrenta}
\eea
the $n$-th components result
\bea \label{atrentuno}
(X_{20})_n &=& {2\over 3} \sqrt{\pi\over 5} 
{i{\o_0}^3 \f_0 \over \o_{_{BD}} + 2} 
\mu a^2 n J_n(ne) \\ 
(X_{2\pm 2})_n &=& \mp 2 \sqrt{\pi\over 30} 
{i{\o_0}^3 \f_0 \over \o_{_{BD}} + 2} \mu a^2 \nn \\ 
&& {1\over n} 
\{(e^2-2) J_n(ne)/(ne^2) + 2 (1-e^2) J_n'(ne)/e \nn \\ 
&\mp& 2(1-e^2)^{1/2} [(1-e^2) J_n(ne)/e^2 - J_n'(ne)/(ne)]\}
\label{atrentadue}
\eea

\section{Appendix}

We want to determine explicitly the form of the scalar GWs radiated
by a bi\-na\-ry system. From (\ref{ventisei}) and
(\ref{aotto}), (\ref{adiciassette}), (\ref{aventiquattro}), 
{\ref{aventicinque}), and taking into account that in the
limit $r\to \infty$ the spherical Hankel functions become
\be \label{buno}
h_l^{(1)} (\o r) \sim {e^{i[\o r-(l+1)\pi/2]}\over \o r}
\ee
one can easily obtain
\bea 
\xi(\vec x, \o) &=& -{2\mu \over (2\o_{_{BD}} + 3)Gr} e^{i\o r} \nn \\
&& \{v^2 + m/d + (\o d)^2/6 +
2i\o (\O_2/m_2 - \O_1/m_1) \hat n\cdot \vec d \nn \\
&& - \o^2 d^2(3{n_z}^2-1)/12 + \o^2[(\hat n\cdot \vec d)^2 -
(n_x d_y - n_y d_x)^2]/4\}_\o
\label{bdue}
\eea
where the subscript $\o$ indicates that all the quantities in the right
member of the above expression (\ref{bdue}) are to be considered
as Fourier components with frequency $\o$ (for example, $v\to v(\o)$).

The time dependent amplitude is \cite{will}
\bea
&& \xi(\vec x, t) = \xi(\vec x, \o) e^{-i\o t} + c. c. =
-{2\mu \over (2\o_{_{BD}} + 3)G r} \nn \\
&& \left[v^2 + {m\over d} -
{1\over 6} {d^2\over dt^2} (d_k d^k) \left(1 - {3{n_z}^2-1\over 2}\right) -
2 {\cal G} \hat n\cdot \vec v -
{d^2\over dt^2} (\hat n\cdot \vec d)^2 
- {(n_x d_y- n_y d_x)^2\over 4} \right]\nn\\
&&= -{2\mu \over (2\o_{_{BD}} + 3)G r}
\left[v^2 + {m\over d} - 2 {\cal G} \hat n\cdot \vec v -
(\hat n\cdot \vec v)^2 + {m\over d^3} \hat n\cdot \vec d \right]
\label{bquattro}
\eea
where we have set $v= v(\o) e^{-i\o (t-r)} + c. c.$, ecc. 
\newpage

\end{document}